\documentclass[11pt]{article}
\usepackage[margin=1in]{geometry}
\usepackage{amsmath,amssymb}
\usepackage{graphicx}
\usepackage{float}
\usepackage{authblk}
\usepackage[colorlinks=true,citecolor=blue,linkcolor=blue,urlcolor=blue]{hyperref}
\usepackage{setspace}
\usepackage{caption}

\title{\Large \textbf{Trend formation with sparse global sampling}}
\author[1]{Sarath Sankar}
\author[2]{Abhijit Sinha}
\author[2]{Shankar Ghosh}
\author[2]{Vikram Tripathi}
\author[2]{Amitava Bhattacharya}

\affil[1]{School of Physics and Astronomy, Tel Aviv University, Tel Aviv 6997801, Israel}
\affil[2]{Tata Instute of Fundamental Research, Homi Bhabha Road, Navy Nagar, Mumbai 400005, India}

\date{\today}

\begin{document}
\maketitle
\thispagestyle{empty}

\bigskip

\noindent\textbf{Abstract}

\noindent
Achieving global coordination without a central controller or dense global communication is a defining challenge for both biological collectives and engineered swarms. We introduce and analyze a minimal model in which self-propelled agents in a bounded domain periodically reorient their motions toward the centroid of a small, randomly chosen subset of their peers, with no direct sensing of any individual neighbor's position or heading. We show that this sparse, non-local sampling rule reliably drives an initially disordered population to a globally aligned, nematic state, and that shrinking the sampled subset -- down to the minimum of two agents -- accelerates ordering rather than impeding it: the resulting estimation noise, filtered through a geometric turning rule, is itself the engine of symmetry breaking. We derive an analytical criterion, in quantitative agreement with simulation, that predicts when this ordering succeeds as a function of the sampling size and sampling frequency.  We confirm the mechanism experimentally in a swarm of up to fifteen differential-drive robots. These results identify sparse random sampling as an information-efficient route to collective coordination, with implications for understanding animal collectives and for designing communication-limited robotic swarms.

\bigskip
\onehalfspacing

\section{Introduction}
Collective behavior is one of the most striking manifestations of emergence in
nature~\cite{marchetti2013hydrodynamics,camazine2001self,vicsek2012collective}.  
Starling murmurations~\cite{ballerini2008interaction,cavagna2010scale}, fish
schools~\cite{katz2011inferring,herbertread2011inferring,tunstrom2013collective},
insect swarms~\cite{attanasi2014finite,buhl2006disorder,puckett2014searching,kelley2013emergent,ni2016tensile},
pedestrian crowds~\cite{helbing1995social,helbing2000simulating}, and vehicular
traffic~\cite{chowdhury2000statistical,tian2017cellular,helbing2001traffic,bando1995dynamical,jiang2001full}
all share a common and arresting feature: coherent large-scale order arising
spontaneously from the interactions of many autonomous agents, none of whom
possesses knowledge of the global state or is necessarily directed
by a central authority~\cite{couzin2009collective,moussaid2009collective,hildenbrandt2010self}. 
The same phenomenology appears in engineered systems such as robotic
collectives~\cite{rubenstein2014programmable,brambilla2013swarm}, distributed
sensor networks~\cite{olfati2007consensus}, autonomous vehicle
platoons~\cite{swaroop1996string}, where coordination must be achieved under 
strict constraints on computation, communication bandwidth, and
energy expenditure~\cite{bonabeau2010swarm}.
The possibility of richly organized collective states emerging from simple rules
has made collective dynamics one of the central paradigms of modern
complex-systems research~\cite{camazine2001self,gros2015self,albano1996self}.

A foundational question in this field concerns the nature and range of
information exchange between agents~\cite{cavagna2014bird} and the decisions that follow information exchange.  
The seminal models of Vicsek~\textit{et al.}~\cite{vicsek1995novel,toner1995long,vicsek2012collective} 
and Reynolds~\cite{reynolds1987flocks}  
demonstrated that local velocity-alignment interactions -- with each agent adjusting
its velocity to match that of its immediate geographical neighborhood -- suffice
to generate flocking and swarming behavior, establishing local interaction rules
as the minimal substrate for collective order. Subsequent work has refined the
notion of ``neighborhood'' considerably. In \textit{metric} models~\cite{vicsek1995novel,dorsogna2006self}, an agent
interacts with all others within a fixed spatial radius; the number of
interaction partners fluctuates with local density, and agents at the periphery
of a sparse group may become effectively isolated. In \textit{topological}
models~\cite{ballerini2008interaction,peshkov2012continuous}, each agent interacts with a fixed number of ``neighbors" (e.g. a fixed number
selected in an increasing order of separation, or deter
mined for example by Voronoi tessellation) irrespective of
their actual physical separation, lending the interaction network structural robustness
against density fluctuations. Visually mediated interactions introduce further
complexity, since occlusion couples the metric and topological descriptions
through the statistics of the local density
field~\cite{strandburg2013visual,pearce2014role}. 
Empirical observations in starling
flocks~\cite{ballerini2008interaction,cavagna2010scale} suggest that topological  
or visual models may be more relevant than purely metric ones.

Despite their success, neighborhood-based models share a structural vulnerability. Because information propagates only through sequences of local contacts, global coherence depends on the connectivity of the interaction network -- a property that can fail catastrophically when the group is sparse, when noise is large, or when density fluctuations create gaps in the contact graph. Under these conditions, neighborhood-sampled systems fragment: subgroups decouple and evolve independently, erasing the global order that had hitherto built up. This fragility is not an artifact of any particular model but a structural consequence of the neighborhood paradigm itself: the very locality that makes these models parsimonious also limits their ability to maintain cohesion at scale. In biological collectives, fragmentation can mean the difference between a school that evades a predator and one that scatters; in robotic systems it can constitute mission failure.

It is therefore natural to consider global communication, in which every agent has access to the states of all others. 
If every agent in a swarm of size $N$ can
access the states of all others, the connectivity of the
interaction network is guaranteed by construction, and
fragmentation becomes impossible regardless of spatial
configuration or noise level. Evidently, the cost of this
global communication has an expensive scaling $\mathcal{O}(N^2)$ per update step, which makes the approach increasingly difficult to scale as the swarm size grows \cite{gustafson2006issues,klavins2004communication}. Biology does not appear to have solved the coordination problem by granting individuals unlimited communication range; if anything, the empirical record suggests the opposite: remarkable global coherence achieved with surprisingly impoverished local information
\cite{ballerini2008interaction}.
This tension, between the global character of the ordered state and the apparent locality of the interactions that produce it, sits at the heart of the present work.

A structure that has become ubiquitous in modern engineered collective systems suggests a resolution. A driver on
a congested road responds in the first instance to the
vehicles in the immediate neighborhood, adjusting speed
and position based on what can be directly observed:
a local, metric interaction indistinguishable in character
from those studied in classical flocking model
\cite{helbing2001traffic}.
Simultaneously, the driver is influenced by a navigation system that provides nonlocal information about traffic conditions across the entire network, allowing an optimal route choice. In this case,
the driver does not need to acquire from the navigation
system the full state of every vehicle on the road, but
instead only a sparse, statistical summary, a kind
of stochastic projection of global information onto a
scalar decision variable. The driver need not know why the recommended route is optimal; the information architecture of the system ensures that acting on this sparse information is, in its macroscopic effect, equivalent to acting on far richer global data \cite{dai2026autonomous,bagabaldo2024impact}.

This observation motivates a question we believe has not been posed in the collective-dynamics literature in its sharpest form: \textit{how sparse can global information be, and still sustain coherent collective order?} The dichotomy between local neighborhood sampling and full global communication leaves a vast intermediate regime unexplored
\cite{jin2020random,boyd2006randomized}. We posit that a qualitatively different
coordination paradigm lies in this regime. Specifically, we
study a model where each agent periodically samples a
small, randomly chosen subset of the full population, uses
this sample to construct a noisy but statistically unbiased
estimate of the global centroid, and adjusts its heading
accordingly. Such a mechanism is global in reach, for any
agent in the population is a candidate for inclusion in
the sample. At the same time, it is sparse in information
cost: the communication overhead scales as $\mathcal{O}(k)$ per
agent per step for a sample of size $k,$ which can be made
sub-linear in $N$ without sacrificing the statistical integrity
of the estimate. The question is whether this sparse
global signal, propagated through independent random
draws rather than through a connected spatial network,
can generate and sustain ordered collective states that
neighborhood-sampled systems cannot maintain.

The problem of achieving global coherence through
nonlocal interactions has a distinguished analogy in a
rather different physical setting: the synchronization of
coupled nonlinear oscillators. In the Kuramoto model~\cite{kuramoto1975self}, each
oscillator adjusts its phase toward the global mean field,
i.e., the centroid of all phase angles on the unit circle, and
the competition between this centroid-seeking coupling
and intrinsic frequency disorder drives a transition from
incoherence to synchronization as the coupling strength
 increases. It is known that replacing the all-to-all adjacency matrix of interactions in the Kuramoto model with a sparse one can lead to consensus formation within fragmented groups and obstruct global phase coherence, see e.g. Ref.~\cite{rodrigues2016kuramoto-fragment}. Our model shares this centroid-seeking structure, but our random sampling scheme (analogous to periodically updating the sparse adjacency matrix) prevents by design the formation of such 
fragments. To the best of our knowledge, the analogous problem of oscillator synchronization with periodic updates of a sparse random adjacency matrix has not been addressed in the literature.

In this paper, we show that sparse global sampling constitutes a distinct coordination paradigm capable of generating robust collective order under severe information constraints. We study a model in which $N\gg 1$ agents confined to a bounded domain move at a constant speed and periodically reorient toward the centroid of a randomly sampled set of $n$ other agents. We show that the interplay between unavoidable sampling fluctuations and centroid-seeking dynamics generates a geometric filtering mechanism that eventually leads to a macroscopic nematic order. The statistical properties of the resulting state are controlled by three parameters: the sampling size, the sampling frequency and the geometry of the bounding domain. Counterintuitively, reducing the amount of information available to each agent accelerates ordering, with the fastest trend formation occurring for the minimal sampling size $n=2$. We derive a predictive criterion for when trend formation succeeds, construct the corresponding phase diagram, and validate the mechanism experimentally using a swarm of differential-drive robots. Together, these results establish sparse global sampling as an information-efficient route to collective coordination and reveal a constructive role for estimation noise in the emergence of collective order.
\section{Model and emergence of nematic order}

Consider a system of \(N\) point particles  in  two-dimensions, moving at a constant speed, $v,$ and updating their directions only at discrete stroboscopic time steps. We are interested in the time evolution of this system of particles. The initial positions
and velocity directions are distributed independently and uniformly,  the former in an area of size $\pi N$ and the latter in $[0,2\pi)$. At each update, every particle randomly samples \(n\) others uniformly  without replacement, and computes their centroid; if the distance to this centroid estimate exceeds a set threshold, \(R_T=\sqrt{N}\), the particle reorients so that its trajectory passes through the estimated centroid, otherwise it continues unchanged. We also define the normalized velocity (inverse time of flight) as $v_{norm}=v/(2R_T).$  We consider two-dimensional (2D) collisionless dynamics which could be regarded as a projection of certain types of three-dimensional (3D) motion onto the 2D plane; for example, particles in 2D that avoid collisions by going slightly off-plane, or alternatively, individually move in 3D at fixed $z$ positions.

\begin{figure}
    \centering
    \includegraphics[width=0.9\columnwidth]{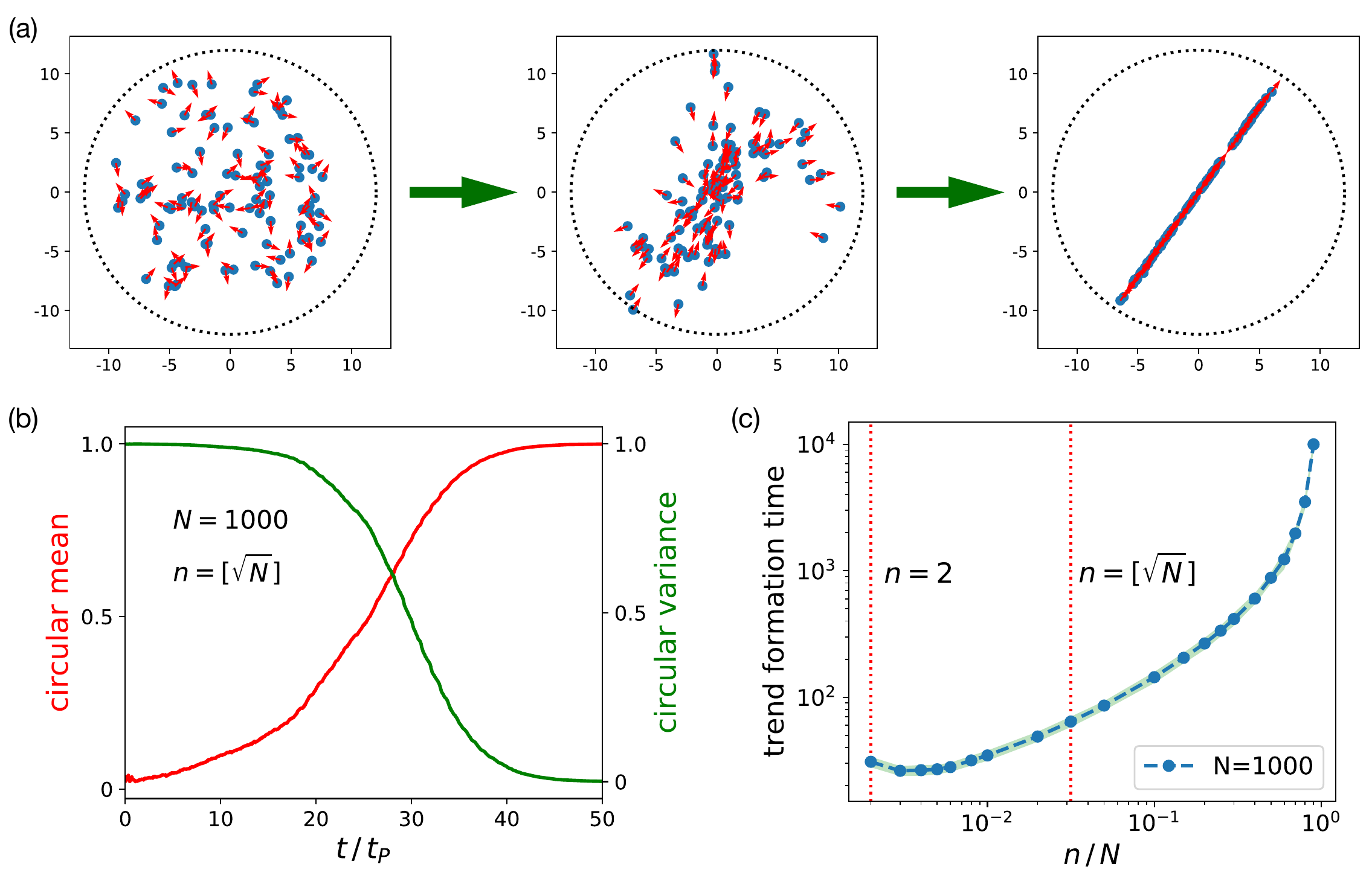}
    \caption{\label{fig:fig1}
    (a) Snapshots of a representative simulation showing the evolution from an initially disordered configuration to a fully aligned state. Arrows indicate particle velocity directions.
    (b) Time evolution of the circular mean (red) and circular variance (green). Here time is shown in units of the ideal passage time, $t_P$.
    (c) Average trend formation time (in units of $t_P$) as a function of the sample size fraction $n/N$, with shaded region indicating standard deviation over realizations.
    }
\end{figure}

Starting from disordered initial configurations, the system robustly evolves toward a dynamical steady state in which all particles move back and forth along a common straight line in $\mathbb{R}^2$.  The orientation of this line is selected spontaneously and is uniformly distributed over $[0,\pi)$, reflecting the rotational symmetry of the domain. Representative snapshots illustrating the evolution from an initially disordered state to a fully aligned configuration are shown in Fig.~\ref{fig:fig1}(a).

We introduce the nematic order parameter (circular mean)
\begin{align}
S(t) & =\left|\frac{1}{N}\sum_{j=1}^{N} e^{i\,2\theta_j(t)}\right|,
\label{eq:oerder_param}
\end{align}
where $\{\theta_j\} \in [0,2\pi)$ are  the velocity angles of the particles. An isotropic angular distribution  of velocities yields $S$ close to $0$, whereas $S$ close to $1$ indicates strong nematic order. To characterize the approach to alignment, we track $S(t)$ together with the circular variance of the velocity angles. As shown in Fig.~\ref{fig:fig1}(b), $S(t)$ grows  from zero while the circular variance decays toward zero, indicating progressive ordering. The time required to reach the aligned state generally increases with sampling size $n$ as shown in Fig.~\ref{fig:fig1}(c), except perhaps for a shallow minimum at $n=3.$ We define the completion of trend formation as the time at which the circular variance falls below a threshold that scales as $1/N.$

\begin{figure}
    \centering
    \includegraphics[width=0.8\columnwidth]{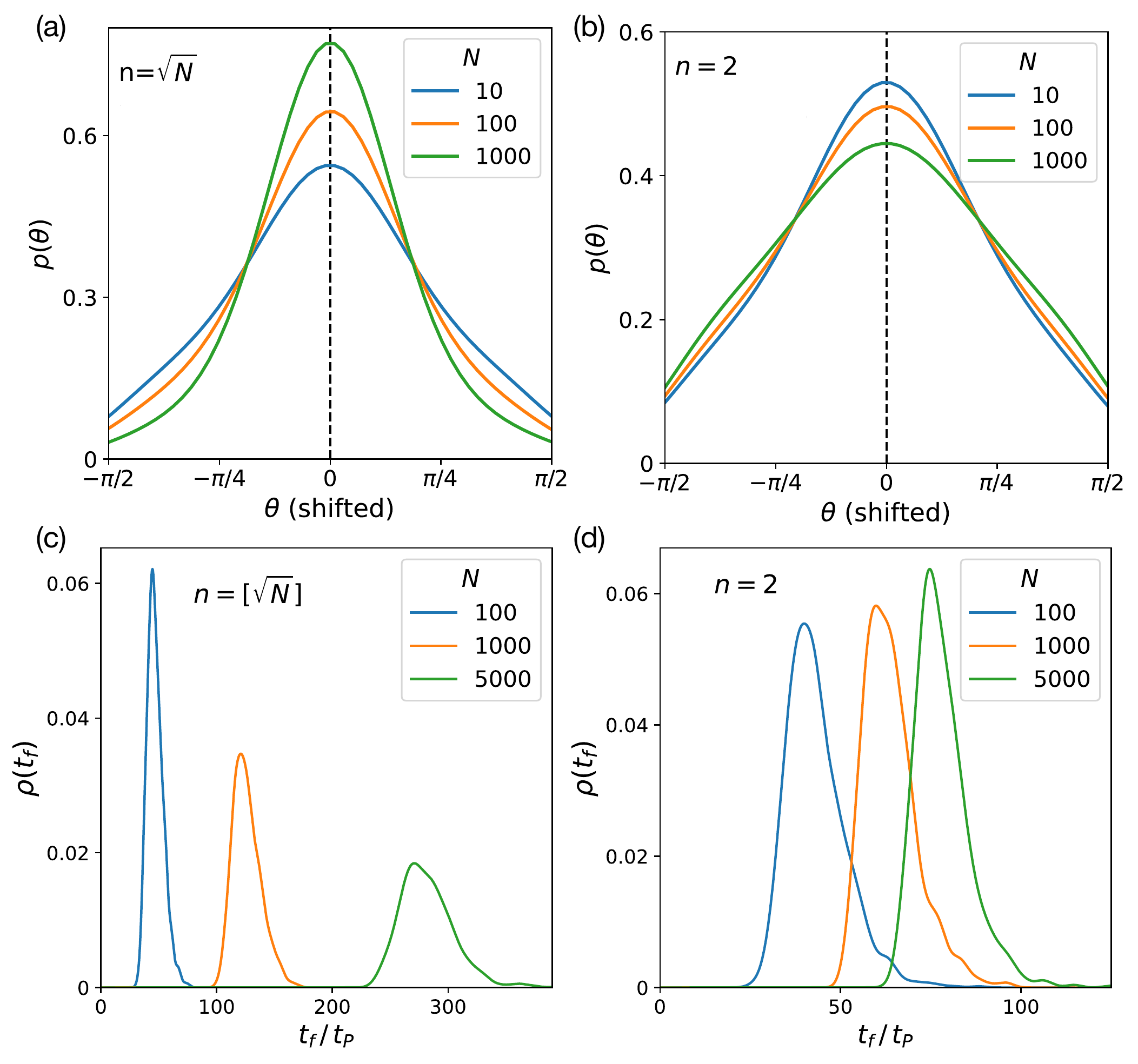}
    \caption{\label{fig:sampling_comp}
    Contrasting trend formation properties between sampling sizes $n=[\sqrt{N}]$ and $n=2$ at $v_{norm}=0.1$. (a)-(b)  Distribution of the final trend angle,$\theta$, starting the evolution from some fixed initial condition. 
    Here the trend angles are shifted by the mean so that the distribution is centered around $\theta=0$. The PDFs are obtained using von-Mises Kernel density estimation. (c)-(d) Distribution of the trend formation time, $t_f$, starting the evolution from random initial conditions chosen from a uniform distribution. The PDFs are obtained using a Gaussian Kernel density estimation.
    }
\end{figure}

While the above establishes global trend formation, it poses an  important question: why does reducing the amount of information available to each agent accelerate the emergence of global order? Since larger samples provide more accurate estimates of the global centroid, one might naively expect them to enhance coordination. The opposite trend observed in Fig. \ref{fig:fig1}(c) suggests that fluctuations in the centroid estimate play a constructive role. This is consistent with a growing body of work showing that fluctuations can sometimes accelerate the emergence of collective order: weak noise has been shown to convert continuous mean-field transitions into sharper, discontinuous ones in self-propelled particle systems \cite{martin2024fluctuation}, and a related paradoxical effect, in which added noise improves the coherence of a collective decision, has been reported in flock-level decision-making models \cite{ferdinandy2012landing}. We now explore the effects of these fluctuations.

\section{Statistical consequences of sparse stroboscopic sampling}

We  examine how sparse stroboscopic sampling influences the statistical properties of the ordered state. Figures \ref{fig:sampling_comp}(a,b) show the distribution of the final trend orientation starting from a given initial configuration. Specifically, we first re-center the distribution corresponding to each initial condition by shifting its mean to $\theta=0$, and then average the re-centered distributions over randomly chosen initial configurations. For sampling size $n=[\sqrt{N}]$, the distribution becomes increasingly sharply localized around its mean value as $N$ grows, whereas for $n=2$ the distribution becomes broader with increasing $N.$ We physically interpret this result as follows. Any typical initial configuration will have weak anisotropies: for example, across any diameter that partitions our circular bounding domain, the number fluctuations around the mean value $N/2$ will be $\mathcal{O}(\sqrt{N}).$ A larger sampling size such as $n=[\sqrt{N}]$ has a higher likelihood of picking up such anisotropies compared to, say, $n=2.$ Thus, although both sampling strategies are ultimately observed to result in an alignment, larger sampling sizes tend to amplify an incipient trend, while for smaller sampling sizes, the amplification of a dynamically generated fluctuation causes trend formation.

The effect of sampling size on the distribution of trend formation times is shown in Figs. \ref{fig:sampling_comp}(c,d). Remarkably, trend formation times for $n=2$ are much shorter than for $n=[\sqrt{N}]$ -- a fact we had also noted in Fig.~\ref{fig:fig1}(c). Moreover, the distribution of trend formation time for $n=2$ progressively becomes narrower with increasing $N,$ which is opposite to the effect seen for $n=\sqrt{N}.$ Physically, this is a consequence of the geometric filtering mechanism (discussed below, see Sec. \ref{sec:trend_formation}) that relies on fluctuations in the estimated centroid position to amplify incipient trends. Increasing the sampling size suppresses fluctuations in the position of the estimated centroid through averaging, thereby weakening the reinforcement mechanism responsible for symmetry breaking. This is consistent with the general understanding that fluctuation effects in collective motion are strongest at small system/sample size and are progressively averaged away as this size grows, recovering mean-field behavior \cite{martin2024fluctuation}.

Considered together, the above results reveal  the following. Smaller-sized samples generate larger centroid fluctuations that accelerate trend formation but produce greater run-to-run variability, whereas larger-sized samples yield more reproducible ordered states at the cost of slower convergence. The fastest route to collective alignment occurs not for the most accurate centroid estimate, but for a noisy one.

 We have identified the role of sampling fluctuations in trend formation, but the existence of fluctuations alone is not sufficient. Successful alignment additionally  is controlled by the particle velocity and the geometry of the confining boundary. We now discuss the geometric filtering mechanism, and based on it, we  derive a criterion for when trend formation would occur.

\section{Trend formation through geometric filtering mechanism} \label{sec:trend_formation}

Each particle estimates the group centroid from a randomly sampled  set of $n$
others (uniformly without replacement), yielding a noisy but unbiased estimate of the true centroid, which, we assume without loss of generality, coincides with the barycenter of the bounding domain. Statistical fluctuations result in initial quadrupolar (nematic) and higher multipole moments apart from a predominantly circular one.
The incipient nematic asymmetry deforms the distribution of the estimated centroid from circular
(Fig.~\ref{fig:trend_vnorm}a) to an elliptical one 
(Fig.~\ref{fig:trend_vnorm}b), which is further amplified by a nonlinear
threshold: a particle at point $P$ which is at a distance $d$ from the true centroid $C$ re-orients its velocity only if its distance $d + X_n$ from the estimated centroid exceeds the distance $R_T$ of the bounding domain boundary. Here $X_n$ is the deviation of the estimated centroid (through an $n$-sized sampling) from $C$ along the line joining the particle and the centroid. Thus for $d < R_T$ only positive noise
fluctuations ($X_n > 0$) trigger reorientation. Combined with the elliptical
geometry, the region biased toward the nematic line subtends a larger area
(Fig.~\ref{fig:trend_vnorm}b, pink) than the region biased away (blue),
driving collective nematic order along the major axis. Crucially, since the major axis of the ellipse tracks the nematic alignment direction, the pink area grows as the order parameter $S$ increases, making the bias self-reinforcing. When $PC$ is perpendicular to the major axis,
the geometric bias vanishes, but  random sampling still produces net
alignment.

The combined action of geometric bias and sampling noise generates a
self-reinforcing alignment mechanism. As order strengthens, transverse motion is
suppressed and the system settles into a nematic, line-aligned state in which
particle headings fluctuate between two antipodal directions along a preferred
axis. We note that $S$ quantifies the degree of alignment to a common axis and
is insensitive to the antiparallel (head--tail) symmetry between these two
directions; whether the system additionally selects a single global direction of
motion (polar order) is not addressed here.

This geometric picture makes a concrete prediction: trend formation should
succeed only when particles turn early enough to repeatedly sample the
asymmetric noise cloud before overshooting the true centroid. Whether this
condition is met depends on the particle speed and sampling size, and it can
fail - as we show next. We now derive a quantitative criterion for when the
geometric filtering mechanism operates successfully.

 \section{Dynamical criterion for trend formation}

\begin{figure}
    \centering
     \includegraphics[width=0.6\columnwidth]{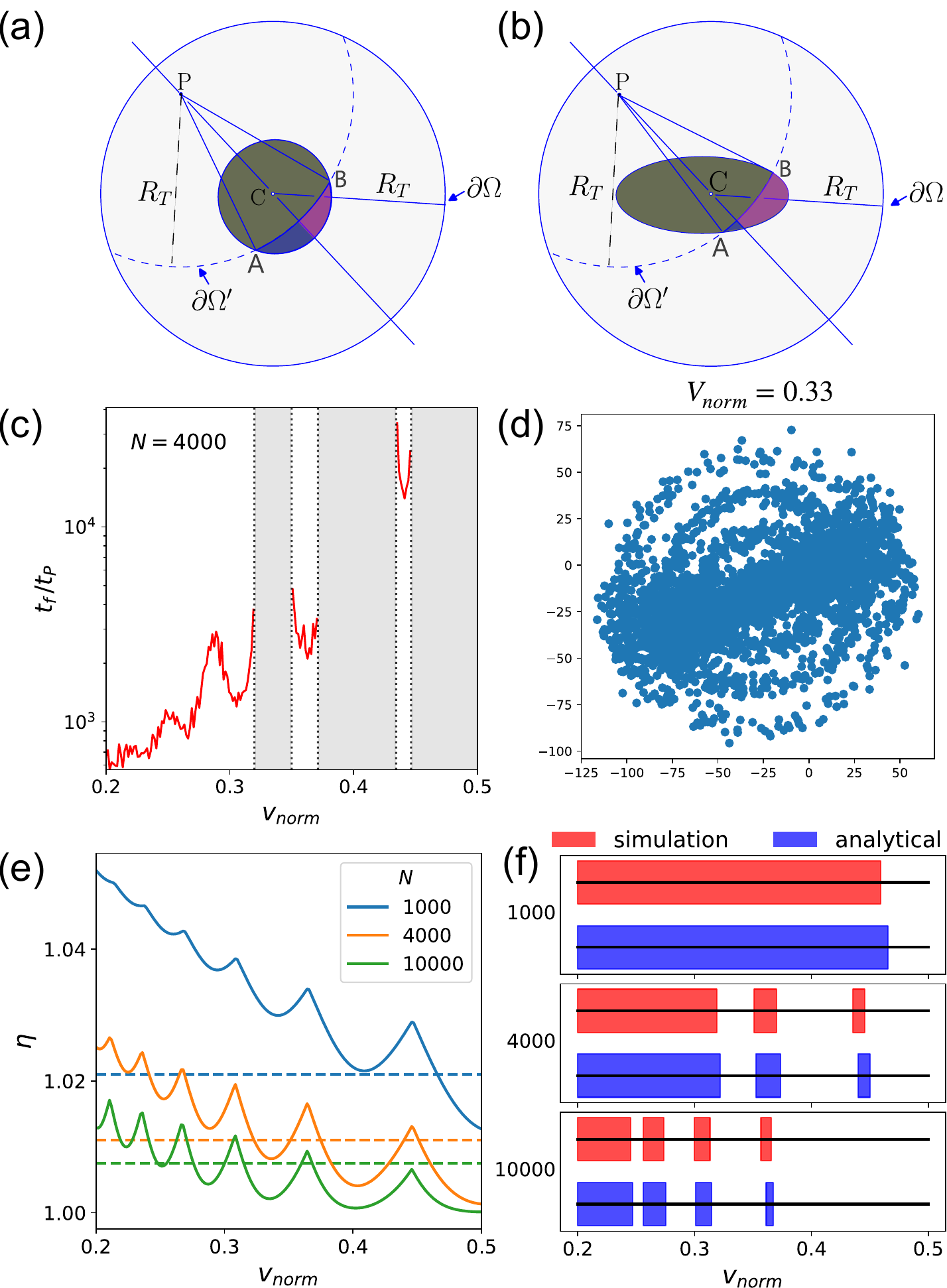}
       \caption{\label{fig:trend_vnorm}   \textbf{Geometric filtering mechanism and dynamical criterion for trend formation}. Agent
$P$, at distance $d$ from the true centroid $C$, infers an estimated centroid through noisy sampling, resulting in a perceived centroid distribution (gray shaded region). Trajectory reversal occurs only when the sampled centroid lies outside $\partial\Omega'$. (a) For an isotropic particle distribution, the noise cloud is circular and no directional bias arises. (b) A weak nematic asymmetry,  either coming from initial configuration or dynamically generated, deforms the noise cloud into an ellipse whose major axis tracks the nematic alignment direction.  Sampling from the pink region returns $P$ toward the nematic line; sampling from the blue region deviates it away. Since the pink area exceeds the blue, and grows further as $S$
 increases,  the ellipse elongates along the alignment direction - reinforcing reorientation events progressively outpace opposing ones, making the bias self-reinforcing.   
       (c) Trend formation time as a function of $v_{norm}$ for the case of $n=[\sqrt{N}]$ sampling with $N=4000$. Trend formation does not occur in the grey regions. (d) An intermediate snapshot of the particle distribution at $v_{norm}=0.33$ where trend formation is frustrated: despite there being a dominant trend direction, a significant fraction of the particles (those that lie on the annular rings) are unable to align since they cannot turn ``early enough" (see main text) . (e) Understanding the trend formation regime analytically by calculating the estimation parameter $\eta$ discussed in the main text. The dashed line denotes $\eta_c$ and trend occurs when $\eta>\eta_c$. (f) The analytically obtained trend formation regimes (colored) are in excellent agreement with our simulations.}
\end{figure}

Figure  \ref{fig:trend_vnorm}(c) shows the trend formation time as a function of the normalized speed $v_{\mathrm{norm}},$ for the case $n=[\sqrt{N}].$
The trend formation time generally increases with $v_{\mathrm{norm}}$; for larger values of the time-step, a particle can overshoot the turning threshold in a single stride, reducing the probability of an early turn and thereby weakening the geometric filtering that drives alignment.
Outside the trend-forming regions, the system fails to reach a globally aligned state.

The origin of this failure is illustrated in Fig.  \ref{fig:trend_vnorm}(d). At sufficiently large values of $v_{\mathrm{norm}}$, some particles cannot reverse their trajectories early enough to join the developing trend. Instead, these particles execute persistent annular orbits around the system, preventing complete alignment. Although a dominant trend direction may still be visible, the coexistence of these non-aligning particles frustrates the globally ordered state. We make this idea more quantitative below.

\subsection{Analytical phase diagram }
The geometric filtering mechanism that drives the trend formation requires a particle to ``turn early" - i.e. the estimated centroid distance when it turns should be greater than the actual centroid distance. We can arrive at the trend-formation phase diagram by checking this condition within the invariant state. In this section we  derive the phase diagram as a function of the velocity (inverse of sampling frequency) for the case, $1 \ll n \ll N$; in particular, we will consider $n=[\sqrt{N}]$. 

Without loss of generality, let us consider the invariant state with  particles moving along the $x$ axis. Each particle will turn at both the left (L) and right (R) ends. Note that the geometric filtering mechanism requires the particle to ``turn early" in at least one end. We define a measure, $\eta$, that captures this ``turning early" and which is easy to calculate:
\begin{equation}
  \eta = \rm{min}_{i}\left(\rm{max}_{\{R,L\}}\left(\eta_i^R,\eta_i^L\right)\right),
  \label{eq:eta_def}
\end{equation}
where,
\begin{equation}
   \eta_{i}^{k=\{L,R\}}= \frac{E(l^k_{\rm{ECM}})}{E(l^k_{\rm{CM}})}.  
\end{equation}
The subscript $i$ denotes the particle index,  $k=L,R$ denotes the turning end, $E(l^k_{\rm{CM}})$ and $E(l^k_{\rm{ECM}})$   denote respectively the expectation value of the actual centroid and conditional expectation value of the estimated centroid when turning (operational definitions for these are discussed below). We now make the following conjecture: trend formation will occur only when,
\begin{equation}
    \eta(v,n) > \eta_c(n).
\end{equation}
Importantly note that  $\eta$ is a function of both $v$ and $n$, but $\eta_c$ is a function of $n$ only. For $\eta<\eta_c,$ a fraction of the particles (such as those in the annular rings in Fig.~\ref{fig:trend_vnorm}(b)) cannot ``turn early" enough to align with any existing trend.

 We now discuss the calculation of $\eta$. Let us rescale distances such that $R_T=1$. Then $v=v_{\rm{norm}}$. To leading order, the particles are uniformly distributed in $[-l,l]$, where $l=E(l_{CM})$, which has to be determined. The variance in the position of a particle is then $l^2/3$. It then directly follows that the variance of (unconditional) estimated centroid is, $l^2/(3n)$. For $n\gg1$, using central limit theorem, the probability distribution of ECM becomes,
\begin{equation}
    \rho(c)=\frac{1}{\sqrt{2\pi}\sigma}\exp{\left[-\frac{c^2}{2\sigma^2}\right]}, \quad {\rm{with}} \quad \sigma=l/\sqrt{3n}.
\end{equation}

Consider a particle moving to the right with some $v_{\rm{norm}} <1$. Suppose the closest position of the particle to $x=0$ is $b$ with $b>0$. The particle would then move along the lattice $b,b+v_{\rm{norm}}+2v_{\rm{norm}}+\ldots$, before it turns back from any of this lattice points. The starting position $b$ is arbitrary. Since we will only consider $v_{\rm{norm}}<1$ we choose $b$ to be uniform in $[0,v_{\rm{norm}}]$. In fact we will do this in a discrete manner: $q$ number of uniformly placed points, $b_0,b_1,\ldots,b_{q-1}$ in $[0,v_{\rm{norm}})$ with uniform weights $w_i=1/q$. We can increase $q$ so that the solution to the system of equations discussed below converges to the desired accuracy.

Consider the case with some $b_i$. Let us denote the position of the particle as, $x_{ij}\equiv b_i + j v_{\rm{norm}}.$
Let $f_{ij}$ denote the probability that the particle turns before $x_{ij}$. We then get the following set of recursion relations,
\begin{equation}
    f_{i,j+1}=f_{i,j} + (1-f_{i,j})h_{i,j+1},
\end{equation}
 with initial condition $f_{i,0}=h_{i,0}$, where,
\begin{equation}
    h_{i,j}=\int_{-\infty}^{x_{ij}-1} dc\, \rho(c) = \frac{1}{2}\left[1+{\rm{Erf}}\left(\frac{x_{ij}-1}{\sqrt{2}\sigma}\right)\right],
\end{equation}
denotes the probability that the ECM, $c$, obeys the turn back condition, $x_{ij}-c > R_T = 1.$ We need only keep $m$ lattice points starting with $b_i$ so that $1-f_{i,m-1}$ is below some desired tolerance. Typically, $m=[2/v_{norm}]$ is good enough. Starting with the initial condition, the recursion equations are easy to solve. We then obtain the probability that a particle would turn at $x_{ij}$,
\begin{equation}
    P_{i,j}=\begin{cases}
        w_i(f_{i,j}-f_{j-1}),\quad j\geq 1\\
        w_i f_0,\quad j=0
    \end{cases}
\end{equation}
The mean turning distance from the actual centroid is then obtained as,
\begin{equation}
    E(l_{CM}^R)=\sum_{i,j}x_{i,j}P_{i,j}.
\end{equation}
Note that $h_{ij}$ depends on $l=E(l_{CM}^R)$ since $\sigma$ depends on $l$; so the solution to the system of equations has to be determined in a self-consistent manner. Typically only a few iterations by using the $l$ from the previous iteration is required to do this. Finally the (conditional) expectation value of the distance of the estimated CM when turning is obtained as,
\begin{equation}
   E(l_{ECM}^R)=\sum_{i,j} l_{i,j} P_{i,j}, 
\end{equation}
where,
\begin{equation}
    l_{i,j}=\frac{1}{h_{i,j}}\int_{-\infty}^{x_{i,j}-1} dc\, (x_{i,j}-c)\rho(c) = x_{i,j}-\frac{c_{i,j}}{h_{i,j}},
\end{equation}
with,
\begin{equation}
    c_{i,j}=\int_{-\infty}^{x_{i,j}-1}dc\, c\,\rho(c)=\frac{-\sigma}{\sqrt{2\pi}}\exp{\left[\frac{-(x_{i,j}-1)^2}{2\sigma^2}\right]}.
\end{equation}
We get a similar set of equations for particles moving to left and by solving it we obtain, $E(l^L_{\rm{CM}})$ and $E(l^L_{\rm{ECM}})$. Then using \eqref{eq:eta_def}, we obtain $\eta$. Fig.~\ref{fig:trend_vnorm}(e) shows $\eta$ as a function of $v_{\rm{norm}}$ for $n=[\sqrt{N}]$. The overall decreasing trend of $\eta$ with $v_{norm}$ is expected: the probability to ``turn early" decreases as sampling frequency decreases.  The oscillatory behavior emerges from the discrete change with $v_{norm}$ of the number of samplings that occur between left and right turns for a particle.  Note that the time to trend formation (Fig.~\ref{fig:trend_vnorm}(c)) depends inversely on $\eta$; in particular the dips in Fig.~\ref{fig:trend_vnorm}(e) become peaks in Fig.~\ref{fig:trend_vnorm}(c). 

The critical values $\eta_c(n)$ need to be determined separately, as they are bridging quantities between the original 2D dynamics and the invariant 1D dynamics that we analyzed above. We determined them empirically from the simulations by demanding agreement with the first no-trend boundary where the trend forming time diverges. $\eta_c(n)$ depends on $n$ implicitly through its dependence on $\sigma(n),$ and in particular we observe that $1-\eta_c(n) \sim 1/(2\pi\sqrt{n}).$ We find a remarkable agreement between the simulation and analytical phase diagrams in Fig.~\ref{fig:trend_vnorm}(f), which also serves as a verification of our conjecture.

\section{Effect of confining domain geometry on trend formation.}  

Within the reinforced stochastic description of alignment, the geometry of the confining domain enters by biasing the initial competition between alignment sectors.  Once a dominant trend direction is established, particle headings fluctuate between two antipodal directions, whose occupations evolve through reinforcement. While the reinforced dynamics governs the amplification and saturation of a trend, the confining geometry determines which alignment sectors are favored and therefore influences both the onset and the orientation of collective order.

\begin{figure}
    \centering
    \includegraphics[width=0.99\columnwidth]{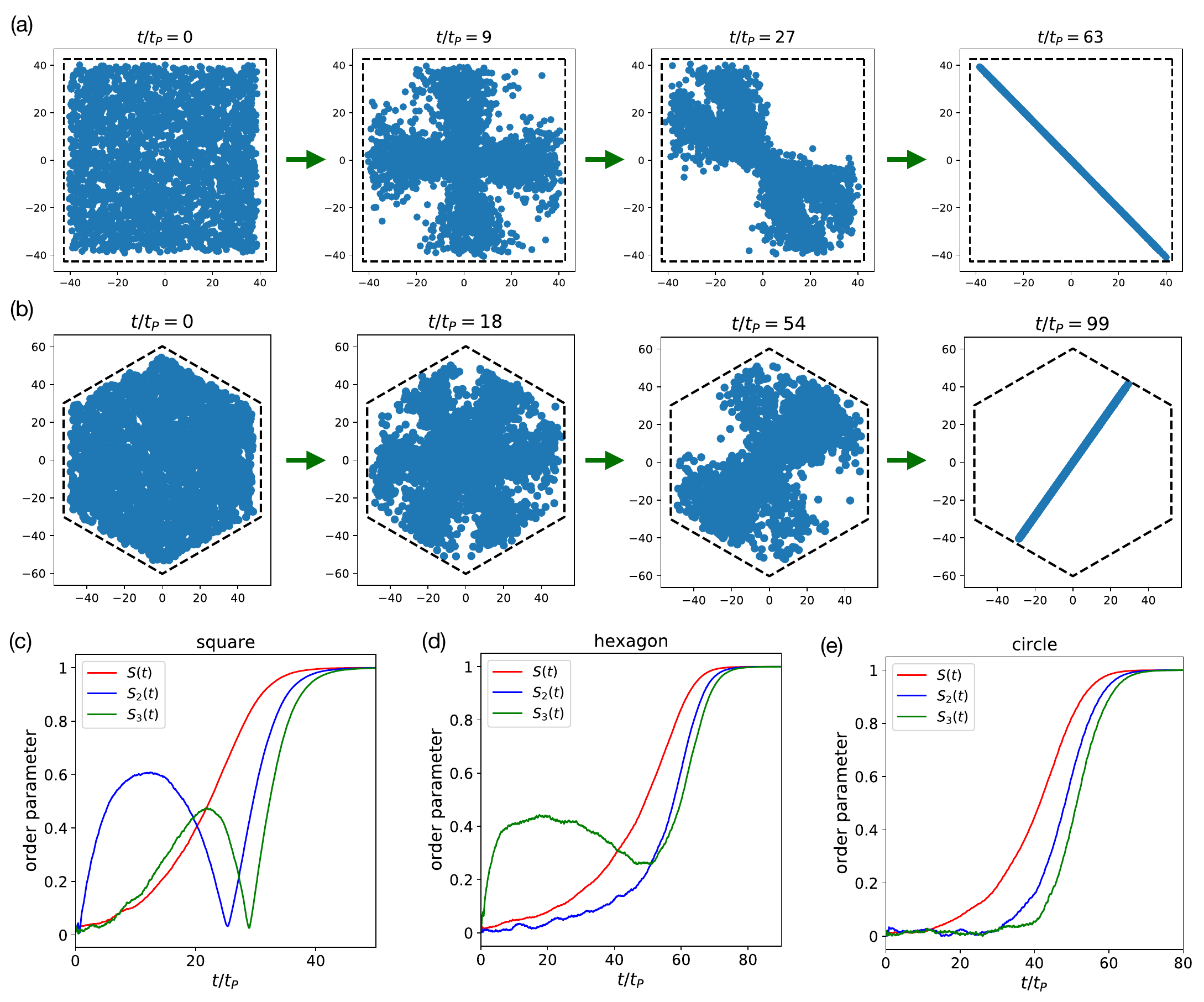}
    \caption{\label{fig:geometry_trend}
    Snapshots of evolution under square and hexagon geometry constraints (the considered polygon orientations are shown with dashed boundaries) are respectively shown in (a) and (b). Here $n=[\sqrt{N}]$ sampling is considered with  $N=3000$ particles and $v_{norm}=0.02$. The associated evolution of the order parameters are shown respectively in (c), (d), and is contrasted with the case of circle in (e).
    }
\end{figure}

For circular domains, the rotational symmetry renders all alignment directions equivalent, which ultimately leads to an arbitrary mean orientation once ordering sets in. In contrast, since (regular) polygon geometries only have discrete rotation symmetries, they provide directional bias. The most striking effect occurs in the early stages of the evolution from a uniform distribution: for $n=[\sqrt{N}]$ sampling, the distribution first reorganizes as a multi-lobed one that is aligned along the normals from the edges to the centroid of the polygon (second column of Fig.~\ref{fig:geometry_trend}(a) and (b)). The underlying mechanism is the directional dependence of early turn events due to the constraining geometry. The onset of such multi-lobed distributions can be captured by higher  order parameters,
\begin{equation}
    S_m(t)=\left|\frac{1}{N}\sum_{j=1}^N e^{i2m\theta_j(t)}\right|.
\end{equation}
Note that when turning back, $\theta \to \theta+\pi$ on the average, and so only even-sided regular polygons can sustain the multi-lobed distribution; for this reason we will consider only even-sided polygons. For the square case $S_2(t)$ is dominant in the early stages of the evolution where as in the hexagon case it is $S_3(t)$ that becomes dominant (see Figs.~\ref{fig:geometry_trend} (c), (d)). This is to be contrasted with the circular domains where $S_1(t)$ stays dominant throughout the evolution (Fig.~\ref{fig:geometry_trend}(e)).

\begin{figure}
    \centering
    \includegraphics[width=0.95\columnwidth]{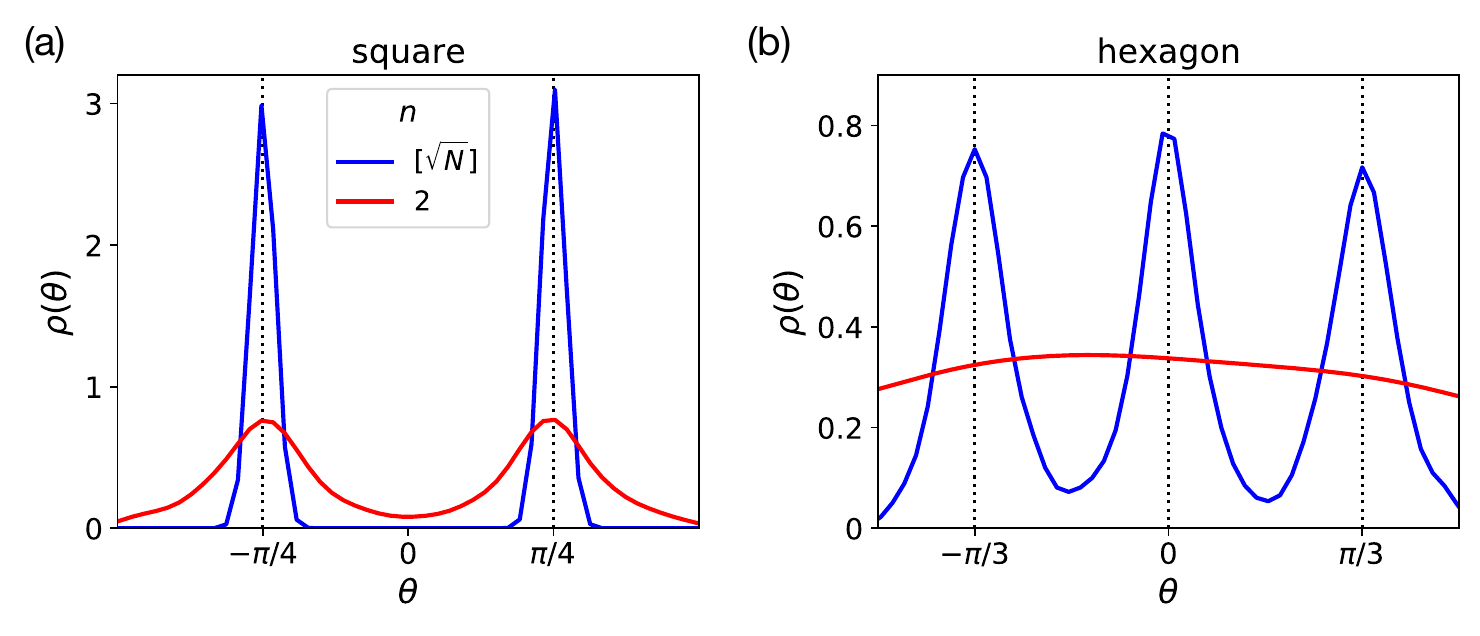}
    \caption{\label{fig:geometry_angle_dist}
    The distribution of the final trend angle for (a) square and (b) hexagon geometry constraints for $n=[\sqrt{N}]$ (blue) and $n=2$ (red) sampling. The polygon orientations are same as that in Fig.~\ref{fig:geometry_trend}. The PDFs are obtained using von-Mises kernel density estimation. 
    }
\end{figure}

Once a dominant $S_m(t)$ is established, the system enters a coarsening regime which is a consequence of our centroid-seeking dynamics whose effect is to make the swarm shape convex and simply-connected. In the square, the angle between the lobes gradually reduces, causing them to merge along a final trend line aligned close to one of the diagonals. In contrast, in the hexagon, the merging occurs into one of the three lobes, yielding a final trend line aligned with one of the edge normals passing through the centroid. We have checked that for a regular polygon with $4p$ edges ($p\geq 1$ is an integer), the final trend favors the diagonals, whereas for the case with $4p+2$ edges, the final trend favors the edge normals passing through the centroid. The statistical distribution of the final trend-line orientation, obtained from simulations initialized with different initial conditions, is shown in Fig.~\ref{fig:geometry_angle_dist} (a), (b). For $n=[\sqrt{N}]$ sampling, the distributions are consistent with the above picture. In contrast, for $n=2$ sampling the distributions become significantly broader. In the square, a broad bimodal distribution remains, with peaks along the diagonals, whereas in the hexagon the distribution becomes nearly uniform. This broadening arises from the large fluctuations in the sampled centroid, which trigger premature turning events and thereby weaken the effect of the bounding geometry constraints on the dynamics. A similar broadening towards a uniform distribution would occur as the sampling frequency is reduced ($v_{\rm{norm}}$ is increased); for, at large $v_{norm},$ the particles would frequently overshoot the bounding geometry and hence the dynamics would closely resemble that of a circular domain.

\section{Robotic swarm experiments}

\begin{figure*}[t]
  \centering
  \includegraphics[width=0.9\linewidth]{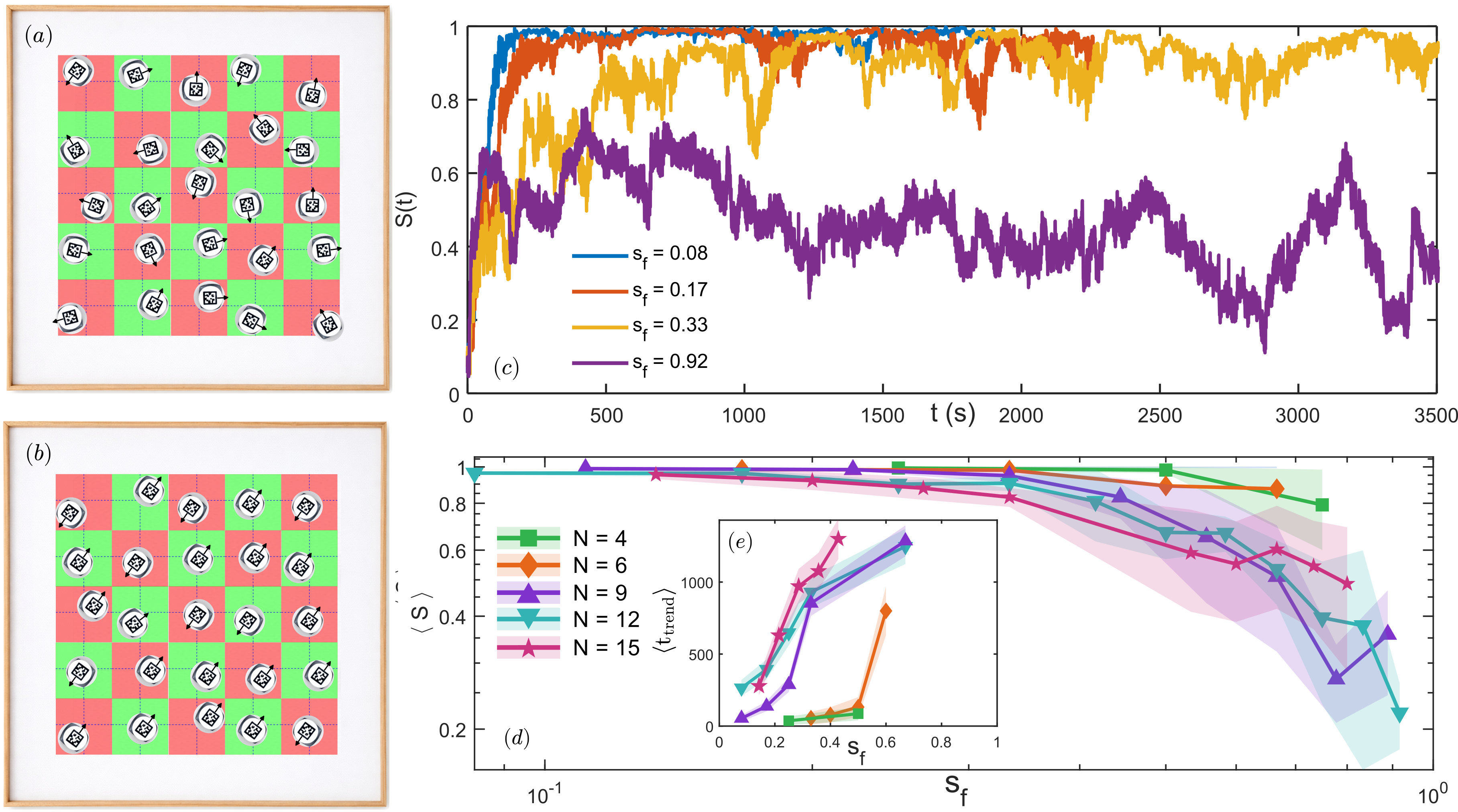}
  \caption{(a,b) Representative image of initial and final configurations of the robot swarm, transitioning from a disordered to a nematically ordered state; squares indicate the fixed offset regions used for collision avoidance.
  (c) Nematic order parameter S(t) vs.\ time $t$ (s) for different sampling fractions $s_f$.
  (d) Steady state mean nematic order $\langle \text{S} \rangle$ vs.\ sampling fraction $s_f$.
    (e) Inset: mean time to reach steady state $\langle t_{\mathrm{trend}} \rangle$ (s) vs.\ sampling fraction $s_f$.}
  \label{fig:Line_Formation_experiment}
\end{figure*}

\paragraph*{\textbf{Experimental Platform and Control}}
The experimental platform consists of aggregates of up to $N=15$ differential-drive robots
operating in a square wooden arena (Fig.~\ref{fig:Line_Formation_experiment}a). Robot
poses $(x,y,\theta)$ are estimated in real time via an overhead Raspberry
Pi camera that detects unique AprilTags at $2\,\mathrm{Hz}$ and streams the
data to a host PC. Onboard PID control, assisted by a current sensor and wheel
encoders, compensates for inter-motor speed variations (see Methods).

\paragraph*{\textbf{Collision Avoidance and Interaction Rules}}
To reconcile finite robot size with a point-particle interaction model, each
robot is assigned a fixed spatial offset from its nominal interaction coordinate.
Collective dynamics are computed in the lab frame as though the robots were ideal
point particles; the result is then mapped to each robot's physical position via
its unique offset. Because offsets are non-overlapping by construction, this
decomposition ensures a minimum physical separation and hence collision
avoidance, without modifying the interaction rules (see Methods). This differs from approaches that enforce collision avoidance by continuously adjusting robot commands in real time to prevent crashes \cite{pickem2017robotarium,wilson2020robotarium}, and is reminiscent of the  virtual-structure approaches to formation control \cite{lewis1997high}, where each agent tracks a fixed offset from an abstract reference frame. However, in our case the offsets are chosen not to enforce any particular formation shape but solely to ensure non-overlap: our construction guarantees safety by design, through a static, non-overlapping assignment of physical offsets, leaving the interaction rules untouched.

Robots are initialized at their offset positions with uniformly random headings
(Fig.~\ref{fig:Line_Formation_experiment}a). At each update, a robot reorients
toward the centroid of $n$ randomly sampled peers (in the lab frame) if its
distance to that centroid exceeds $R_T = \sqrt{N}$, within an angular tolerance
of $\pm 5^{\circ}.$ Between updates each robot travels in a straight line.
\subsection{Results and observations}
Starting from the disordered initial state (Fig.~\ref{fig:Line_Formation_experiment}a),
the robots are observed to align and form a nematic phase
(Fig.~\ref{fig:Line_Formation_experiment}b). The experiments
(Fig.~\ref{fig:Line_Formation_experiment}c,d) broadly reproduce the simulations,
though real-world factors introduce quantitative differences. In simulations,
$S\approx 1$ for all values of the sampling fraction $s_{f}=n/N:$ in experiments, however, $\langle S
\rangle$ decreases monotonically with $s_{f}$, a discrepancy we attribute to
imperfect reorientation updates driven by mechanical noise. Crucially, the two regimes
converge in the low-$s_{f}$ limit -- arguably the most physically interesting
regime, where each agent processes the least information. That robust and stable nematic order emerges here demonstrates that dense
sampling is not a prerequisite for such a collective alignment, but rather sparse, local information
is sufficient. The mean time $\langle t_{\text{trend}}\rangle$ to reach the steady state grows qualitatively with $s_{f}$, but finite-size fluctuations scaling as
$1/\sqrt{N}$ elevate the noise floor, making it difficult to resolve the trend formation time $\langle
t_f\rangle$ precisely at larger sampling fractions
(Fig.~\ref{fig:Line_Formation_experiment}c,e).

\section{Discussion}

We have shown that global alignment in an active particle system localized in a convex bounding domain can be
induced by stochastic sampling in the strongest measure-zero limit:
a sample of just $n=2$ agents. That individual updates involving just two randomly
drawn agents suffice to coordinate the behavior of a population of arbitrary size is  perhaps the most surprising result of this paper. We identified the
geometric mechanism underlying the trend formation which is the result of the interplay of subsystem sampling and centroid seeking dynamics. For small sampling sizes, the final orientation is not very sensitive to higher multipole fluctuations in the initial velocity distribution and relies more on amplification of some dynamically generated fluctuation. For large sampling fractions, small initial asymmetries in the velocity distribution bias the noise
cloud of the estimated centroid, generating a self-reinforcing drift toward
the trend formation along the leading initial quadrupolar moment of the velocity distribution. Remarkably, a sampling size $n=2$ is not merely sufficient but achieves global order faster
than any larger sample size, for the variance of the centroid estimate at
small $n$ amplifies the reinforcement mechanism rather than suppressing it.  

Our mechanism for trend formation driven by centroid-seeking sampling is reminiscent of the Kuramoto synchronization  under conditions of sparse connectivity ~\cite{rodrigues2016kuramoto-fragment}, which, however, is known to be susceptible to fragmentation. In our case, the effect of dynamical updates of the connectivity matrix mitigates the fragmentation tendency. The framework presented here also points toward a richer landscape of collective phases
accessible through modifications of the information structure. For example, we have observed that introduction of a
constant directional bias in the selection of the sampled centroid results in a transition from the linear nematic state to a stable rotating
annular configuration, with the inner radius of the annulus apparently tunable
by the magnitude of the bias angle. While a detailed understanding of such transitions is beyond the scope of the present work, their existence
suggests that geometric and informational constraints on the sampling process
can act as control parameters for engineering qualitatively distinct collective
phases -- a direction we regard as a promising avenue for future investigation. 

More broadly, our results establish that global coordination is achievable with the absolute minimum of information. This is not a
computational approximation to full communication but a distinct coordination
paradigm, one in which the stochastic geometry of extreme undersampling is
itself the engine of order. These properties make sparse global subsampling a
natural candidate mechanism for collective organization in biological systems
operating under severe metabolic and sensory constraints, and a  design
principle for distributed robotic systems where bandwidth and energy are
limited. The  mapping to reinforced stochastic processes developed
here may further serve as a general tool for studying many-body systems in
which agents act on compressed statistical summaries of a global state rather
than on direct local observations.

\section*{MATERIALS AND METHODS}

\subsection*{Simulation model}

We simulated $N$ point particles moving at constant speed $v$ in a two-dimensional circular domain of area $\pi N$, so that the number density remains $O(1)$ for large $N$. Initial positions and velocity directions were drawn independently and uniformly. Headings were updated synchronously at discrete stroboscopic time steps: at each update, particle $i$ sampled a subset $\mathcal{S}_i$ of $n$ other particles uniformly without replacement and computed the estimated centroid $\vec{R}_i^{\mathrm{ECM}} = \frac{1}{n}\sum_{j \in \mathcal{S}_i} \vec{r}_j(t)$. If $\|\vec{R}_i^{\mathrm{ECM}} - \vec{r}_i(t)\|$ exceeded the threshold $R_T = \sqrt{N}$ (the domain radius), the particle's heading was updated so that its velocity direction intersected $\vec{R}_i^{\mathrm{ECM}}$; otherwise the particle continued unchanged. Between updates, particles moved ballistically. The normalized speed $v_{\mathrm{norm}} = v/(2R_T)$, the sampling size $n$, and the shape of the confining domain (circle, or a regular $4p$- or $(4p+2)$-sided polygon) were varied systematically as described in the main text; order parameters $S(t)$ and $S_m(t)$ were computed from the resulting trajectories, and results were averaged over multiple independent realizations for each parameter set.

\subsection*{Robotic platform and experimental arena}

Experiments were conducted with aggregates of up to $N=15$ differential-drive robots confined to a bounded square planar arena. Arena dimensions were scaled with system size to maintain a fixed number density across all experiments, following the general design of overhead-tracked swarm testbeds such as the Robotarium~\cite{pickem2017robotarium}, e-puck~\cite{mondada2009epuck}, and Pheeno~\cite{wilson2016pheeno}. Robot poses $(x,y,\theta)$ were estimated in real time by a calibrated overhead camera detecting AprilTag markers affixed to each robot, yielding time-stamped position and heading data at $2\,\mathrm{Hz}$. Onboard PID control, assisted by wheel encoders and a current sensor, compensated for inter-motor speed variations. Sensing, computation, and data logging were handled centrally by a host PC in wireless communication with the robots.

\subsection*{Implementation of sparse-information interactions}

Although computation was centralized for synchronization and logging, the interaction rules were information-sparse by design. At each discrete update step, the host PC emulated decentralized agent behavior by restricting each robot's accessible information to a randomly drawn subset of $k$ peers, sampled uniformly without replacement from the remaining $N-1$ robots. All heading updates depended solely on the robot's own state and the states of its sampled subset, faithfully implementing the sparse-sampling dynamics of the theoretical model.

\subsection*{Collision avoidance via frame decomposition}

To reconcile the point-particle assumption of the theoretical model with the finite physical extent of the robots, we used a frame-decomposition strategy. Each robot was assigned a fixed spatial offset from its nominal interaction coordinate; interactions and centroid estimates were computed in the abstract lab frame, treating robots as point particles, and the resulting heading commands were then mapped back to each robot's physical position via its unique offset. Because offsets were chosen to be non-overlapping, this mapping guaranteed a minimum inter-robot separation at all times, ensuring collision avoidance by construction without altering the effective interaction rules.

\subsection*{Interaction protocol}

Each experimental trial comprised two sequential phases. In the initialization phase, robots were guided by a closed-loop position controller to predetermined home positions with uniformly randomized headings, producing reproducible initial conditions across trials. In the interaction phase, the stochastic sampling dynamics was activated: at each update, every robot estimated the centroid of its sampled peers in the lab frame, and if its distance to this estimated centroid exceeded the threshold $R_T$, it reoriented its heading toward the centroid within an angular tolerance of $\pm 5^{\circ}$; otherwise it maintained its current heading. Between updates, robots moved ballistically.

\subsection*{Parameter exploration}

Experiments were performed across a range of system sizes $N$ and sampling fractions $s_f = n/N$, with multiple independent trials per parameter set to ensure statistical robustness. The update interval and distance threshold $R_T$ were held fixed across all experiments and chosen to place the system firmly within the dynamical regime described by the theoretical model.

\section*{Acknowledgments}
 We thank S. Gupta (TIFR) for a useful discussion on sparse Kuramoto models. S.S. acknowledges TIFR for hosting him while a part of this work was completed. A.S., S.G, V.T. and A.B. acknowledge the support of the Department of Atomic Energy, Government of India, under Project Identification No.\ RTI 4002, and TIFR for providing computational resources.


\end{document}